\documentclass[superscriptaddress,secnumarabic,amssymb,amsmath,nobibnotes,aps,prd,nofootinbib,preprint]{revtex4-2}

\usepackage{mathrsfs}
\usepackage{graphicx}
\usepackage{latexsym}
\usepackage{amsfonts}
\usepackage[dvips]{epsfig}
\usepackage{color}

\begin{document}

\title{Thermostatistical analysis and negative heat capacities of Yukawa and Lee-Wick potentials in noncommutative phase spaces}


\author{Maria G.  Sousa}\email{maria.girlandia@estudante.ufjf.br}
\affiliation{Department of Physics, Universidade Federal de Juiz de Fora, MG, Brazil}

\author{Everton M. C. Abreu}\email{evertonabreu@ufrrj.br}
\affiliation{Department of Physics, Universidade Federal Rural do Rio de Janeiro,  RJ, Brazil}
\affiliation{Applied Physics Graduate Program, Physics Institute, Universidade Federal do Rio de Janeiro, RJ, Brazil}

\author{Albert C. R. Mendes}\email{albertcrm@ufjf.br}
\affiliation{Department of Physics, Universidade Federal de Juiz de Fora, MG, Brazil}

\author{M. J. Neves}\email{mariojr@ufrrj.br}
\affiliation{Department of Physics, Universidade Federal Rural do Rio de Janeiro, RJ, Brazil}



\begin{abstract}
\noindent In recent years, physical models based on noncommutative algebras have attracted 
considerable interest, as they provide a natural framework to incorporate a 
fundamental scale, often associated with semiclassical aspects of quantum gravity. 
Noncommutative geometry modifies the underlying phase-space structure, potentially 
leading to new insights into unresolved problems in theoretical physics. In this work, we adopt a semiclassical approach to perform a thermostatistical 
analysis of well-established interaction models, namely the Yukawa and Lee--Wick 
potentials, within a noncommutative phase space. We investigate how phase-space 
deformations affect the density of states, partition function, mean energy, and 
heat capacity, considering both microcanonical and canonical ensembles within the 
Boltzmann--Gibbs framework. Our results show that the introduction of the noncommutative parameter $\Theta$ 
induces nontrivial modifications in thermodynamic quantities, including qualitative 
changes in the heat capacity. In particular, regions with negative heat capacity 
may emerge, which we interpret as signatures of the limitations of the semiclassical 
and perturbative treatment rather than definitive physical effects. The analysis is carried out under the assumption of weak noncommutativity and 
$|\beta V(r)| \ll 1$, which constrains the regime of validity of the results. 
Within this domain, our findings highlight the role of phase-space geometry in 
shaping thermodynamic behavior.
\

\end{abstract}

\date{\today}




\maketitle

\newpage

\section{Introduction}

The noncommutative (NC) phase-space theories have emerged as a promising framework in theoretical physics, challenging conventional conceptions of spacetime continuity and offering new insights into the microscopic structure of matter and interactions. These approaches are motivated by several independent indications, ranging from fundamental theories of gravity to condensed matter systems under extreme conditions, in which spacetime itself may exhibit a discrete or deformed structure at very short distances.

NC frameworks have also been extensively explored in other areas of theoretical physics, such as high-energy and quantum field theories, where divergences naturally arise. The introduction of a NC algebra in these contexts aims to mitigate such divergences by imposing a fundamental length scale in spacetime, thereby naturally regularizing ultraviolet (UV) divergences \cite{1,1.1,2.1,2.3,2.4,2.5,2.6,2.7,2.8,2.9,2.10,2.11,2.12,2.13,2.14,2.15,2.16,3,3b,4,5}. At very short distances, Lorentz invariance may be effectively violated; however, in the limit where the NC parameter vanishes, the conventional commutative results are smoothly recovered \cite{1,1.1,5,6,7,8,9,10,10.1}.

NC phase-space formulations offer a complementary alternative to conventional renormalization-based approaches, particularly in view of their known limitations \cite{4,Szabo}. By preserving scale invariance, these frameworks provide a unified description across energy scales, from the ultraviolet (UV) to the infrared (IR) \cite{Minwalla,2.3}. As a result, they enable the construction of models whose predictions can, in principle, be confronted with experimental or observational data over a broad range of regimes.

Recent studies have explored the implications of NC structures in a variety of contexts, including modified uncertainty relations, quantum thermodynamics, and thermal machines. In these settings, noncommutativity has been shown to affect energy spectra, entropy production, and the efficiency of thermodynamic cycles, leading to potentially observable deviations from standard predictions \cite{refA,refB,refE,refG}.

While these approaches are typically formulated within fully quantum or operator-based frameworks, the present work adopts a complementary perspective based on phase-space noncommutativity. This allows for a direct investigation of how geometric deformations influence the density of states and the resulting thermostatistical behavior of interacting systems.

In contrast to previous studies, which mainly focus on single-particle or quantum thermodynamic setups, our analysis addresses interacting systems governed by short-range potentials, providing a bridge between NC geometry and semiclassical statistical mechanics.

Motivated by these considerations, this work aims to investigate how deformations of the phase-space structure induced by noncommutativity affect the thermodynamic behavior of interacting systems. In particular, we focus on two important models of short-range interactions: the Yukawa and Lee--Wick potentials, which are relevant in different physical contexts, such as nuclear, plasma, and effective field theories \cite{Y1,Y2,LW1,LW2,LW3}.

The central question is whether NC corrections can induce qualitative changes in thermodynamic properties, such as the emergence of nontrivial regimes or anomalous behavior, even in systems that are not inherently long-range. This question is further motivated by the broader perspective that thermodynamic properties may encode information about the underlying geometry of phase space, a viewpoint that has gained attention in different approaches to emergent gravity.

Within this context, exploring the thermodynamics of systems with deformed phase-space structures may provide insight into how geometric modifications at the microscopic level influence macroscopic observables and possibly mimic characteristics typically associated with strongly correlated or long-range interacting systems. Furthermore, an additional motivation for investigating the thermodynamics of short-range potentials lies in the connection between finite-range interactions and emergent phenomena in gravity and cosmology.

In recent years, works such as those of Jacobson \cite{Jacobson} and Verlinde \cite{Verlinde} have explored the idea that gravity may emerge as an entropic phenomenon, linked to the thermodynamics of horizons and holographic principles. In this context, a detailed understanding of how short-range potentials influence thermodynamic quantities, such as entropy, temperature, and heat capacity, may provide insights into the microscopic origin of gravitational dynamics and the thermodynamic description of confined systems, including stars, atomic nuclei, and cosmological models with effective short-range interactions \cite{15,landsberg,nss,bms,nicolini}.

This work is structured as follows. In Section 2, we present the NC extensions of the Yukawa and Lee--Wick potentials, detailing the derivation of the modified Hamiltonians, equations of motion, and conserved quantities. Section 3 is devoted to the microcanonical analysis, where we compute the density of states and the temperature for the Yukawa and Lee--Wick potentials in the NC framework. In Section 4, we perform the canonical ensemble analysis, obtaining the partition functions, mean energies, and heat capacities with explicit dependence on the NC parameter. Finally, in Section 5, we present the conclusions and perspectives for future work.

We use natural units $\hbar = c = 1$ throughout this paper, in which the NC $\Theta$-parameter has dimensions of area.

\section{Non-commutative  classical model of  Yukawa and Lee-Wick potential}

In NC theory, a particular realization of these frameworks is characterized by the commutation relation between spatial coordinate operators, {\it i.e.}, $[\,\hat{x}_i \, , \, \hat{x}_j \, ] = i\hbar\, \tilde{\Theta}_{ij}$, where $\tilde{\Theta}_{ij}$ is a real antisymmetric matrix defining the noncommutativity parameters. Such NC spaces have found numerous applications, including studies of black holes and massive gravity in general relativity, effective models in condensed matter physics, and various formulations in string theory \cite{10,10.1}, among others.

In the strictly classical case, one can introduce a deformation by modifying the usual symplectic structure, generating an algebra $\mathscr{A}_{\alpha}$ equipped with the star product \cite{Djemai}
\begin{eqnarray}
(f * g)(x) = \exp\!\left(\frac{i}{2}\,\alpha_{ab}\,\partial^a \partial^b\right) f(x) \, g(x') \Big|_{x'=x} \; ,
\end{eqnarray}
where $f$ and $g$ are differentiable functions or fields, and $\alpha_{ab}$ defines the deformed symplectic structure given by
\begin{equation}
\label{0.1}
\alpha_{ab} = \begin{pmatrix}
\tilde{\Theta}_{ij} &  \delta_{ij} \\
-\delta_{ij} & 0
\end{pmatrix} \; .
\end{equation}
The spatial coordinates $x^{i}$ and the linear momentum components $p^{i}$ satisfy the Poisson algebra
\begin{eqnarray}\label{relxipi}
\{ \, x_i \, , \, x_j \, \} = \tilde{\Theta}_{ij} \, , \quad \{ \, x_i \, , \, p_j \, \} = \delta_{ij} \, , \quad \{\, p_i \, , \, p_j \, \} = 0 \; ,
\end{eqnarray}
where $\tilde{\Theta}_{ij}$ is an antisymmetric matrix containing the NC parameter, with dimensions of area in natural units. It can be written in terms of the Levi--Civita tensor as $\tilde{\Theta}_{ij} = \varepsilon_{ijk} \, \Theta^k$. This relation will be useful throughout this work. In the Hamiltonian formalism, $\Theta^k$ is the NC parameter that appears in the equations of motion \cite{11,12}.

In this context, we consider the Hamiltonian $H$ describing a system of two particles interacting via a central potential $V(r)$, given by
\begin{eqnarray}\label{2}
H=\frac{p_{i}\,p^{i}}{2\mu}+V(r) \; ,
\end{eqnarray}
where $\mu$ is the reduced mass of the system and $p_{i}$ denotes the linear momentum.

Using the Poisson brackets from Eq.~(\ref{relxipi}), and considering Eq.~(\ref{2}) for a central potential $V(r)$, we obtain the system of Hamilton equations
\begin{eqnarray}\label{3}
\dot{x}_{i}=\frac{p_{i}}{\mu}+\epsilon_{ijk}\,\frac{1}{r}\frac{\partial V(r)}{\partial r}\,\Theta_{j}\,x_{k} 
\; , \; 
p_{i}=\mu \, \dot{x}_{i}+\mu \, \epsilon_{i j k} \, \Omega_{j} \, x_{k} \, ,
\end{eqnarray}
where $\Omega_{j}=\frac{1}{r} \frac{\partial V(r)}{\partial r} \Theta_{j}$ plays the role of an effective angular velocity. These equations describe the velocity and momentum of a particle as observed in a non-inertial NC frame with angular velocity $\Omega_{j}$ \cite{11,12}.

From these relations, we obtain the equations of motion corresponding to a modified Newton's second law
\begin{eqnarray}\label{4}
\mu \, \ddot{x}_{i}=-\frac{x_{i}}{r} \frac{\partial V(r)}{\partial r}+\mu \, \epsilon_{i j k} \, \dot{x}_{j} \, \Omega_{k}+\mu \, \epsilon_{i j k} \, x_{j} \, \dot{\Omega}_{k} \; .
\end{eqnarray}
In this way, corrections to Newton’s second law arise that depend on the NC parameter and on the variation of the central potential. These corrections can be interpreted as perturbative effects induced by the NC structure \cite{11,12}.

Although the Hamiltonian in Eq.~(\ref{2}) is a constant of motion, the components of the angular momentum $L_i = \mu\, \epsilon_{ijk}\, x_j\, \dot{x}_k$ are not conserved, and $L_i$ does not generate rotations in the NC framework. However, the generator of rotations around the $\Theta$-axis is conserved. Therefore, restricting to $\Theta_{i}=(0,0,\Theta)$ \cite{11,12}, we obtain
\begin{equation}
\label{2.3}
L_{\Theta} = \Theta\left[ \, (x p_y - y p_x) - \Theta\, \mu\, V(r) - \frac{\Theta\, p_z^2}{2} + \Theta\, \mu\, H \, \right] \; .
\end{equation}

Since the Hamiltonian is conserved, we can identify the following constant of motion:
\begin{equation}
\label{2.4}    
M = x p_y - y p_x - \Theta\, \mu\, V(r) - \frac{\Theta\, p_z^2}{2} \; ,
\end{equation}
which generalizes the canonical angular momentum component $L_z = x p_y - y p_x$ in the presence of the noncommutativity.

The additional terms proportional to the NC parameter $\Theta$ arise from the deformation of the phase-space structure and encode corrections to the standard orbital dynamics. In this sense, $M$ can be interpreted as a deformed conserved quantity associated with rotational symmetry in the NC framework.

Throughout this work, $M$ is treated as a fixed parameter characterizing a given dynamical configuration of the system. Consequently, the thermodynamic analysis is performed at fixed $M$, without introducing an additional statistical averaging over this quantity.

This conserved quantity plays a central role in the physical characterization of the system, particularly in determining how NC corrections influence the thermodynamic behavior of the potentials considered.

Applying this formulation to the Yukawa and Lee--Wick potentials, we now obtain their corresponding NC corrections in spherical coordinates.

\subsection{The NC Yukawa potential}

The well-known Yukawa potential is given by
\begin{equation} \label{2.5}
V_{YP}(r) = -\,k \, \frac{e^{-\mu r}}{r} \; ,
\end{equation}
where $k$ is the coupling constant and $\mu$ is an effective parameter related to the reduced mass of the two-body system, which defines the characteristic interaction length scale. Since the Yukawa potential is central, we introduce spherical coordinates $(r,\theta,\phi)$, and the Hamiltonian in Eq.~(\ref{2}) takes the form
\begin{equation} \label{2.6}
H= \frac{\mu}{2} \, {\dot r}^2 + \frac{\mu}{2} \, r^2 \,{\dot{\phi}}^2 + \mu \, r^2 \, \dot{\phi} \, \Omega + V_{YP}(r) \; ,
\end{equation}
where the angular velocity $\Omega$, and the constant of motion $M$ are given by
\begin{equation}
    \label{2.7}
    \Omega = -\left( \frac{\mu}{r}+ \frac{1}{r^2}\right) \Theta \, V_{YP}(r) 
    \; , \;
    M = \mu r^2 \, \dot{\phi} - \left( \, \mu^2 r+ 2\mu  \,\right)\Theta \, V_{YP}(r) \; .
\end{equation}
From Eq.~(\ref{2.7}), eliminating $\dot{\phi}$ in the Hamiltonian, we obtain
\begin{equation}\label{2.8}
H= \frac{\mu}{2}\,{\dot r}^2 + \frac{M^2}{2\mu r^2} + V_{NCYP}(r) \; ,
\end{equation}
where $V_{NCYP}(r)$ denotes the noncommutative Yukawa potential
\begin{equation} \label{2.9}
V_{NCYP}(r)= -\frac{k e^{-\mu r}}{r}  - \frac{\Theta \, k \, M \, e^{-\mu r}}{r^3} \; .
\end{equation}
Noncommutativity introduces a correction to the Yukawa potential that scales as $r^{-3}$.
\begin{figure*}[th]
\centering
\includegraphics[width=0.49\linewidth]{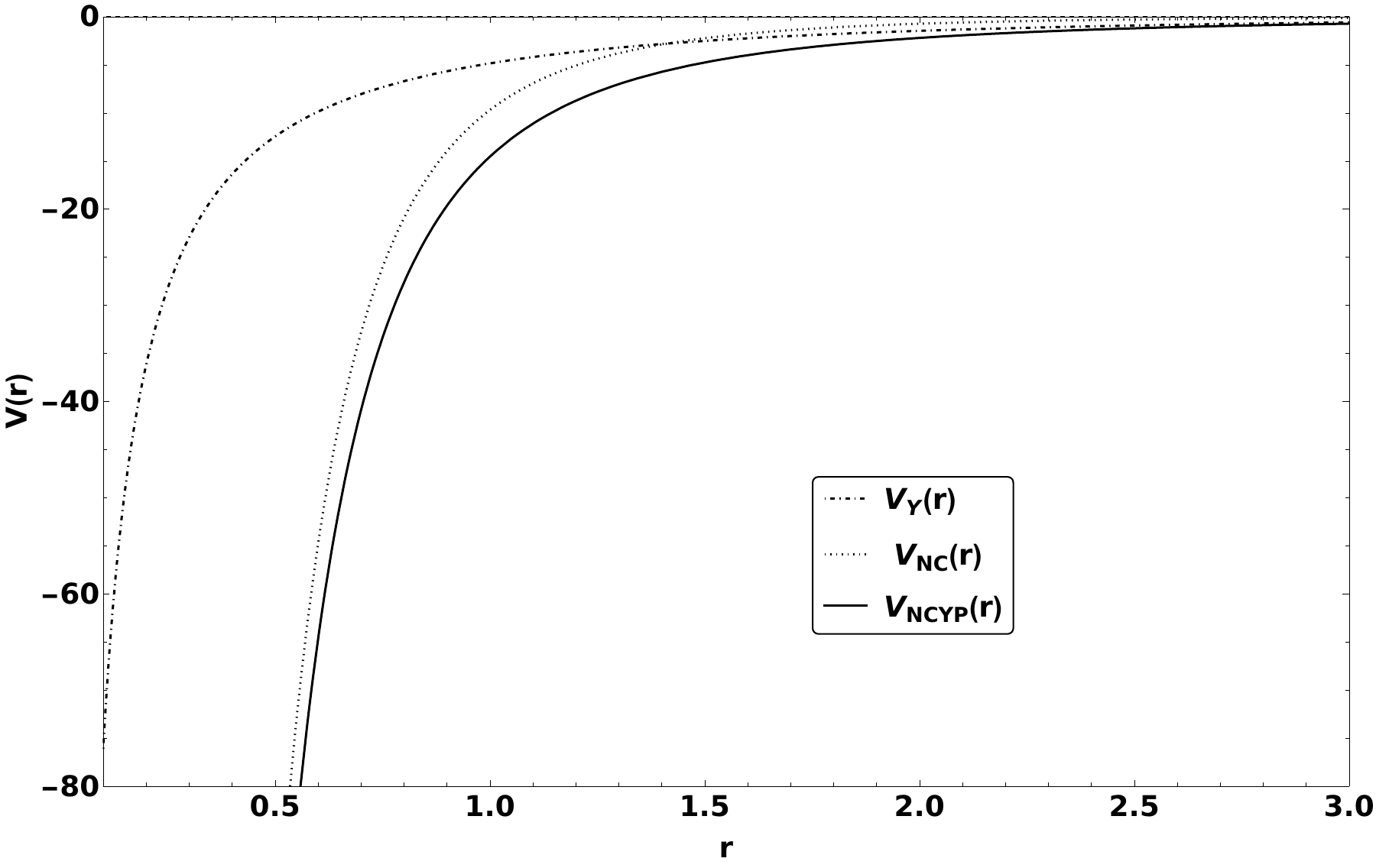}
\includegraphics[width=0.49\linewidth]{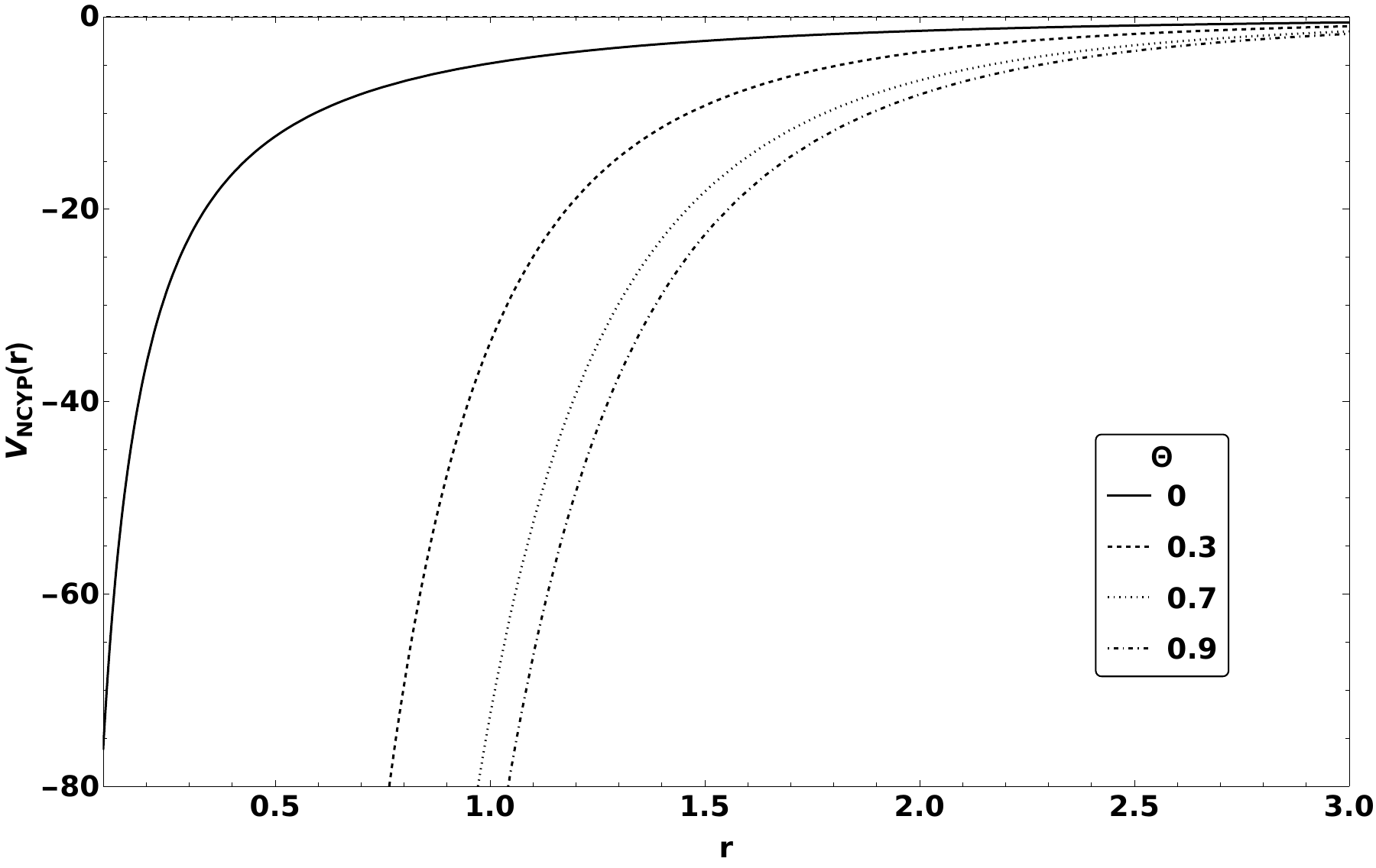}    
\caption{Left panel: Comparison between the standard Yukawa potential $V_Y(r)$, the noncommutative correction term, and the resulting noncommutative potential $V_{NCYP}(r)$ as functions of the radial distance $r$. Right panel: Noncommutative Yukawa potential $V_{NCYP}(r)$ for $\Theta=0.3$, $\Theta=0.7$, and $\Theta=0.9$ (in units of area).}
\label{figYNC1}
\end{figure*}

The NC Yukawa potential is illustrated in Fig.~\ref{figYNC1}. The left panel shows the comparison between the standard Yukawa potential and its NC correction. The right panel displays the resulting NC Yukawa potential for different values of $\Theta$.

\subsection{The NC Lee--Wick potential}

The Lee-Wick (LW) potential is given by
\begin{equation}\label{2.10}
V_{LWP}(r)= k \, \frac{1- e^{-\mu r}}{r} \; .
\end{equation}
This potential can be interpreted as the difference between the Coulomb and Yukawa potentials, and it remains finite at the origin, {\it i.e.}, $V_{LWP}(0)=\mu\,k$.

The Hamiltonian in spherical coordinates has the same structure as in Eq.~(\ref{2.6}) and it is read as
\begin{equation}\label{2.11}
H= \frac{\mu}{2}{\dot r}^2 + \frac{\mu}{2} \, r^2 \, {\dot{\phi}}^2 + \mu \, r^2 \, \dot{\phi} \, \Omega + V_{LWP}(r) \; ,
\end{equation}
where the angular velocity $\Omega$, and the constant of motion $M$ are, respectively,
\begin{equation}\label{2.12}
\Omega = -\mu \, k \, \Theta \, \frac{e^{-\mu r}}{r^2} -\frac{\Theta}{r^2} \, V_{LWP}(r) 
\;\,\, \mbox{and} \;\,\, 
M = \mu r^2 \dot{\phi} + \mu^2 \, k \, \Theta \, e^{-\mu r} - 2\mu \, \Theta \, V_{LWP}(r) \; .
\end{equation}
Eliminating $\dot\phi$ from Eq.~(\ref{2.12}) in the Hamiltonian (\ref{2.11}), we obtain
\begin{equation}
    \label{2.13}
     H= \frac{\mu}{2}{\dot r}^2 + \frac{M^2}{2\mu r^2} + V_{NCLWP}(r) \; ,
\end{equation}
where $V_{NCLWP}(r)$ denotes the noncommutative Lee--Wick potential:
\begin{equation}
    \label{2.14}
    V_{NCLWP}(r) = \frac{k}{r} \, ( 1 - e^{-\mu r}) + \frac{ M \, k \, \Theta}{r^3} \, ( 1 - e^{-\mu r}) \; .
\end{equation}
\begin{figure*}[th]
\centering
\includegraphics[width=0.49\linewidth]{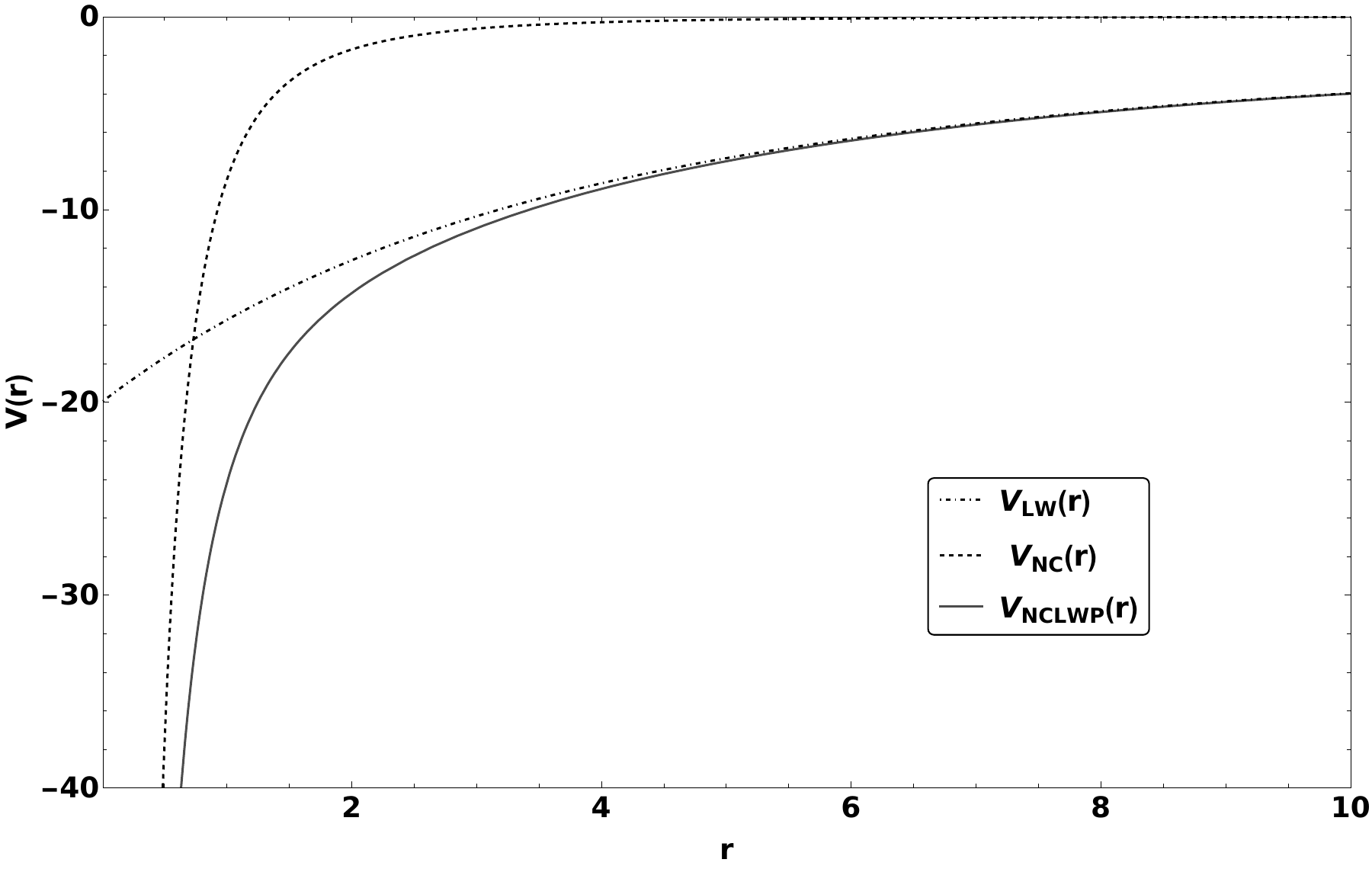}
\includegraphics[width=0.49\linewidth]{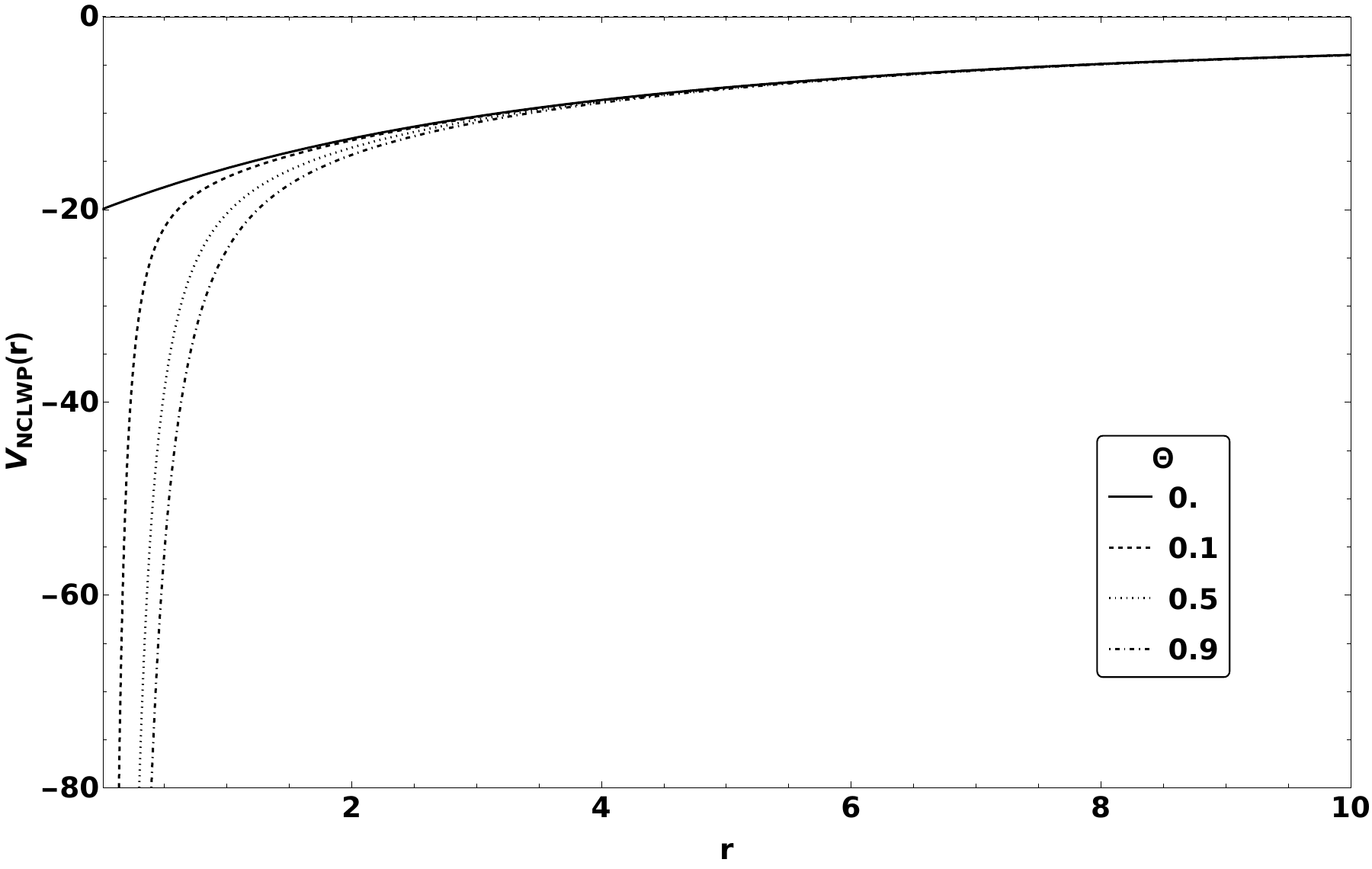}
\caption{Left panel: Comparison between the standard Lee--Wick potential $V_{LW}(r)$, the noncommutative correction, and the resulting potential $V_{NCLWP}(r)$. Right panel: Noncommutative Lee--Wick potential for $\Theta=0$, $0.1$, $0.5$, and $0.9$ (in units of area).}
\label{figLWNC1}
\end{figure*}

The NC contribution in Eq.~(\ref{2.14}) scales as $r^{-3}$, leading to a divergence of the potential at the origin when $\Theta \neq 0$. Figure~\ref{figLWNC1} illustrates this behavior. The left panel shows the comparison between the standard LW potential and the NC correction, while the right panel displays the resulting potential for different values of $\Theta$, including the commutative case $\Theta=0$, which remains finite at the origin.

\section{Thermodynamic Properties of Yukawa and Lee-Wick Potentials and negative heat capacity in NC systems}
In this section, we establish the thermostatistical framework  for short-range interaction potentials within a NC phase space. Our goal is to investigate how the introduction of the NC parameter $\Theta$ modifies the thermodynamic properties of systems governed by Yukawa and Lee–Wick interactions.
The Yukawa and Lee–Wick potentials were chosen because they describe physically relevant short-range interactions across different domains. The Yukawa potential models nuclear and screened Coulomb interactions; the Lee–Wick potential emerges in higher-derivative field theories, and effective interactions in condensed matter physics. These models are therefore representative of distinct physical mechanisms involving screening and exponential decay of interactions, where NC corrections could induce measurable deviations.

For systems with short-range interactions, it is common to employ the microcanonical 
description. In the thermodynamic limit, and under conditions of extensivity and weak 
correlations, this description is typically equivalent to the canonical ensemble. 
However, in finite or confined systems, such as the one considered here, this 
equivalence is not guaranteed.

In particular, the introduction of NC phase-space deformations may 
modify the density of states and affect the effective structure of correlations. 
As a result, the conditions under which ensemble equivalence holds should be carefully 
reassessed in the present framework, especially in light of known situations where 
ensemble inequivalence may arise \cite{paper5}. In this context, the use of the 
microcanonical formalism is motivated by its direct connection with the density of 
states, while remaining consistent with the expected behavior in regimes where 
standard assumptions apply.

In Boltzmann statistical mechanics, for short-range interactions and under appropriate 
conditions, both energy and entropy are additive with respect to subsystems.

\subsection{Canonical and Microcanonical Ensembles}
When we are working with the canonical ensemble, it means that we are operating at a fixed temperature to analyze other thermodynamic properties of the system, such as the internal energy. On the other hand, when considering the microcanonical ensemble, we are fixing the energy of the system to analyze the density of states, temperature, and the heat capacity.
To construct the statistical description of the system under analysis, we will start with the construction of the microcanonical ensemble. Considering the system represented by the Hamiltonian constructed in the position and momentum coordinate phase space.
In the microcanonical ensemble, the energy of the system, $E=H$, is a constant, meaning that it remains fixed. From this ensemble, we can obtain the thermostatistical properties of the system, such as the density of states, {\it i.e.}, $g(E)$, see \cite{12.1}, which is relevant for calculating the entropy, temperature.
Therefore, considering $E=H$ in this formalism, the density of states is defined by 
\begin{eqnarray}\label{46}
g(E)=\frac{1}{N!} \, \int\delta\left[E-H({\bf r},{\bf p})\right] \, d^3{\bf r} \, d^3{\bf p} \; ,
\end{eqnarray}
where $N! \, g(E)$ corresponds to the volume of the phase space occupied by the constant energy surface itself \cite{15}. As we said above, eq. (\ref{2}), this system consists of two particles, with different or equal masses, interacting via $V(r)$. We will presume that the quantity $r$ fluctuates in the interval $(b,R)$. It is tantamount to suppose that the particles are hard spheres of radius $b/2$. The system is confined to a spherical box of radius $R$. By integrating this equation, we can rewrite the expression for the density of states as follows
\begin{eqnarray}\label{47}
g(E)=\text{AR}^3 \int_{b}^{r_{\max }} r^2 \left[ \, E-V(r) \, \right]^2 \, dr \; ,
\end{eqnarray}
where $A = 64\pi^5\mu^3/3$, $R$ is the radius of the confined spherical system, where the particles interact through the central potential $V(r)$, $\mu$ is the reduced mass, and $b$ is the radius of the particles.  The upper limit $r_{\max}$ is determined by the condition $V(r_{\max}) = E$, considering the geometric constraint $b \, < \, r_{max} \, < \, R$.
The density of states is related to the entropy and temperature of the system, respectively, as follows
\begin{equation}
\label{48}
S(E) = \ln{g(E)}\,\,\,\,\text{and}\,\,\,\, \frac{1}{T(E)} = \frac{\partial S(E)}{\partial(E)} \; ,
\end{equation}
where all the interesting thermodynamics proprieties of the system can be understood from the $T(E)$ curve \cite{12.1,12.2}. 

We will use just below  this formalism for the density of states to find the general expression for the density of states and temperature of Yukawa and Lee-Wick potentials in NC. After that, we will calculate the partition function and the heat capacity for this system.

\subsection{Thermodynamic properties for NC Yukawa potential}
%

%

The density of states for this case is given by eq. \eqref{47}
\begin{eqnarray}\label{50}
g_{NCYP}(E)=\text{AR}^3 \int_{b}^{r_{\max }} r^2 \, \left[\,E-V_{NCYP}(r)\,\right]^2 \, dr \; .
\end{eqnarray}
where $V_{NCYP}(r)$ is the NC Yukawa potential eq.(\ref{2.9}).
Therefore, considering terms up to first order in $\Theta$-parameter, the density of states $g(E)$ is read as
\begin{equation}\label{51.1}
g_{NCYP}(E)= \text{AR}^3 \int_{b}^{r_{\max }} r^2\left[ E^2 + \frac{2E k e^{-\mu r}}{r} + \frac{2E \Theta k M e^{-\mu r}}{r^3}   + \frac{k^2 e^{-2\mu r}}{r^2} + \frac{2\Theta k^2 M e^{-2\mu r}}{r^4} \right] \; .
\end{equation}
Evaluating the integrals in eq.(\ref{51.1}), we obtain up to first order in $\Theta$, the following result 
\begin{equation}
    \label{51.2}
    g_{NCYP}(E)=g_{YP}^{(0)}(E) +\Theta \, g_{YP}^{(1)}(E) \; ,
\end{equation}
where $g_{YP}^{(0)}(E)$ is the density of states for the commutative case \cite{15}
\begin{eqnarray}   
\label{52}
g_{YP}^{(0)}(E) &= &\frac{A R^3 E^2}{3} \left\{r_{\text{max}}^3 - b^3 -\frac{3k^2}{2E^2 \mu}\left(e^{-2\mu r_{\text{max}}}-e^{-2\mu b}\right) \right.\nonumber\\  &-& \left.\frac{6k}{E\mu}\left[e^{-\mu r_{\text{max}}}\left(\frac{1}{\mu} + r_{\text{max}}\right) -e^{-\mu b}\left(\frac{1}{\mu} + b\right )\right]\right\} \; ,
\end{eqnarray}
and $g_{YP}^{(1)}(E)$ is the correction for the NC term of the density of states, given by
\begin{eqnarray}
    \label{51.3}
    g_{YP}^{(1)}(E) &=& 2AR^3 k M\Bigg\{
    E\left[ \, \mbox{Ei}(\mu b) - \mbox{Ei}(\mu r_{\text{max}}) \, \right]  \nonumber\\  &+& \ k\left[ \, \frac{e^{-2\mu b}}{b} - \frac{e^{-2\mu r_{\text{max}}}}{r_{\text{max}}} + 2\mu\left(\mbox{Ei}(2\mu b) - \mbox{Ei}(2\mu r_{\text{max}})\right) \, \right] \Bigg\} \; ,
\end{eqnarray}
where $\mbox{Ei}$ is the exponential integral function, $r_{\text{max}}$ is the maximum radial distance, determinate by the condition $V_{NCYP}(r=r_{\text{max}}) = E$, that explicitly, is
\begin{eqnarray}
\label{52.1}
 &&r_{\text{max}} \approx -\frac{k}{E - k\mu} 
- \frac{M E \Theta}{k} \; , \   \text{for }\   -\frac{k e^{-\mu b}}{b}  - \frac{\Theta \, k \, M \, e^{-\mu b}}{b^3}<\frac{E}{k}<-\frac{k e^{-\mu R}}{R}  - \frac{\Theta \, k \, M \, e^{-\mu R}}{R^3},\nonumber \\
\end{eqnarray}
and
\begin{eqnarray}
&& r_{\text{max}}=R \; , \ \  \text{for}  \ \  \ -\frac{k e^{-\mu R}}{R}  - \frac{\Theta \, k \, M \, e^{-\mu R}}{R^3}<\frac{E}{k}<\infty \; .
\end{eqnarray}

%
%
The result eq.\eqref{52} shows how the introduction of the NC $\Theta$-parameter modifies the thermostatistical behavior of the system in comparison with the traditional case ($\Theta=0$), in which the usual result presented in \cite{15} is recovered. 
Therefore, the inverse of the temperature in terms of energy for NCYP to first order in $\Theta$ is read as
\begin{equation}
    \label{52.3}
    \frac{1}{T_{NCYP}(E)} = \frac{1}{g_{NCYP}(E)}\frac{\partial g_{NCYP}(E)}{\partial E} = \frac{1}{g_{NCYP}{(E)}}\left\{\frac{\partial g_{YP}^{(0)}(E)}{\partial E} + \Theta\frac{\partial g_{YP}^{(1)}(E)}{\partial E} \right\} \; ,
\end{equation}
where
\begin{eqnarray}\label{53}
\frac{1}{T_{NCYP}(E)} &=& \frac{1}{g_{NCYP}(E)} \left\{ 
\frac{2g_{YP}^{(0)}(E)}{E}
    - \frac{AR^{3}E}{k }
    \left\{
        B(E)
        + E r_{\max}^{4}
        \left(
            1 - \frac{6 e^{-\mu r_{\max}}}{E r_{\max}}
        \right) \right.\right.  \nonumber \\
        &-& \left.\left. \frac{k^{3}}{E^{2}}
        \left[
            \left(
                1 - \frac{E r_{\max}^{2}}{k}
            \right)
            e^{-2\mu r_{\max}}
            - e^{-2\mu b}
        \right]
    \right\} \right. \nonumber \\ &+& \left. 2AR^3 k M \Theta \left\{
    - \frac{E^2}{2k^2} \left[ r_{\max} + \frac{k}{E} e^{-\mu r_{\max}} \right]^{2}
    + \left[ \operatorname{Ei}(\mu b) - \operatorname{Ei}(\mu r_{\max}) \right] \right. \right.
    \nonumber\\ &+& \left. \left. \frac{k}{(k\mu - E)^{2}} C(E)
    \right\}
\right\} \; ,
\end{eqnarray}
with the functions $B(E)$ and $C(E)$ defined as
\begin{equation}
    \label{53.1}
     B(E) = \frac{6k}{E\mu} \left[ e^{-\mu r_{\max}} \left( r_{\max} + \frac{1}{\mu} \right) - e^{-\mu b} \left( b + \frac{1}{\mu} \right) \right] \; ,
\end{equation}
\begin{equation}
    \label{53.2}
    C(E) = - E\frac{e^{\mu r_{\max}}}{r_{\max}}
    + k \left[
        \frac{e^{-2 \mu r_{\max}}}{r_{\max}^2}
        + 2 \mu \frac{e^{-2 \mu r_{\max}}}{r_{\max}}
        - 2 \mu \frac{e^{2 \mu r_{\max}}}{r_{\max}}
    \right] \; .
\end{equation} 
%
%
Analyzing the eqs. \eqref{52} and \eqref{53} and comparing it with the results eqs. (38) and (40) presented in the \cite{15}, the main difference between is in the additional terms due to the inclusion of the $\Theta$-parameter. In other words, in the semiclassical regime analysis for the NC Yukawa potential, there are relevant corrections to the value of the density of states and the inverse of the temperature. The usual result presented in \cite{15} is recovered when $\Theta =0$.

\subsection{Thermodynamic properties for NC LW potential}

%

%

The density of states to the Lee-Wick potential in NC phase space  is defined by
\begin{eqnarray}\label{56}
g_{NCLWP}(E) = AR^3 \int_{b}^{r_{\max }} r^2 \, \left[E - V_{NCLWP}(r)\right]^2 \; dr \; ,
\end{eqnarray}
in which we use the NC LW potential $V_{NCLWP}(r)$ from the eq. (\ref{2.14}). Considering terms up to first order in $\Theta$, the result of eq.(\ref{56}) is
\begin{equation}
    \label{56.1}
    g_{NCLWP}(E)=g_{LWP}^{(0)}(E) +\Theta \, g_{LWP}^{(1)}(E) \; ,
\end{equation}
where \(g_{LWP}^{(0)}(E)\) is the density of states for the commutative case \cite{15}
\begin{eqnarray}   
\label{56.2}
g_{LWP}^{(0)}(E) &= &-\frac{2AR^3 k^2 E}{\mu}\left\{ \frac{\mu k}{6E^2} \left[ \left( 1-\frac{E\,r_{\text{max}}}{k}\right)^3 - \left( 1-\frac{E\,b}{k}\right)^3 \right] \right. \nonumber\\ &-& \left. e^{-\mu r_{\text{max}}}\left[ \frac{1}{E}\left( 1 -\frac{e^{-\mu r_{\text{max}}}}{4}  \right)-\frac{1}{k}\left( r_{\text{max}} + \frac{1}{\mu}\right)\right] \right. \nonumber \\ &-& \left.  e^{-\mu b}\left[ \frac{1}{E}\left( 1 -\frac{e^{-\mu b}}{4}  \right)-\frac{1}{k}\left( b + \frac{1}{\mu}\right)\right] \right\} \; ,
\end{eqnarray}
and $g_{LWP}^{(1)}(E)$ is the correction for the NC term of the density of states, given by
\begin{eqnarray}
    \label{56.3}
    g_{LWP}^{(1)}(E) &=&-2AR^3M\left\{Ek\left[log\left(\frac{r_{\text{max}}}{b}\right) -\left[\mbox{Ei}(-\mu r_{\text{max}}) -\mbox{Ei}(-\mu b)\right]\right] \right. \nonumber\\ &+& \left. k^2 \left[\frac{1}{r_{\text{max}}}\left[1-e^{-\mu r_{\text{max}}}(2-e^{-\mu r_{\text{max}}})\right] + 2\mu\left[Ei(-\mu r_{\text{max}}) -Ei(-2\mu r_{\text{max}})\right] \right.\right. \nonumber\\  &-& \left.\left.\frac{1}{b}\left[1-e^{-\mu b}(2-e^{-\mu b})\right] + 2\mu\left[\mbox{Ei}(-\mu b) -\mbox{Ei}(-2\mu b)\right] \right]\right\} \; ,
\end{eqnarray} 
where the maximum radial distance $r_{\text{max}}$ is  determinate by the condition $V_{NCLWP}(r=r_{\text{max}}) = E$, that explicitly, is 
\begin{eqnarray}
 &&r_\text{{max}} \approx \frac{\sqrt{2}}{\mu}\sqrt{\mu -\frac{E}{k}} - \frac{M\Theta \left[\mu -\frac{\mu}{2}\sqrt{\mu -\frac{E}{k}} \right]}{2\left(\mu -\frac{E}{k}\right)\left( \frac{\sqrt{2}}{\mu}\sqrt{\mu -\frac{E}{k}} - \frac{3}{2}\right)}\; , \ \ \ \text{for}\nonumber \\  &&\frac{k}{b} \, ( 1 - e^{-\mu b}) + \frac{ M \, k \, \Theta}{b^3} \, ( 1 - e^{-\mu b})<\frac{E}{k}<\frac{k}{R} \, ( 1 - e^{-\mu R}) + \frac{ M \, k \, \Theta}{R^3} \, ( 1 - e^{-\mu R}) \, ,
 \end{eqnarray}
 and
 \begin{eqnarray}
&& r_\text{{max}}=R \; , \ \ \  \text{for}\ \  \ \frac{k}{R} \, ( 1 - e^{-\mu R}) + \frac{ M \, k \, \Theta}{R^3} \, ( 1 - e^{-\mu R})<\frac{E}{k}<\infty \; . 
\end{eqnarray}
The eq. \eqref{56.1} shows that the NC space modifies the functional structure of the phase space, introducing explicit $\Theta$-dependent corrections to the thermodynamic quantities. Although the analysis is developed to first order in $\Theta$, these corrections change the analytical form of the expressions, rather than merely providing small numerical shifts. The logarithmic and exponential-integral terms appearing in the solution reflect the cumulative effects of short-range interactions. When $\Theta = 0$, our result reduces to the classical density of states associated with a short-range potential \cite{15}. For $\Theta \neq 0$, even a small semiclassical NC correction noticeably alters the behavior of the density of states. In particular, the expression for $r_{\text{max}}$ acquires a $\Theta$-dependent modification, which affects the integration limits in the definition of $g_{\text{NCLW}}(E)$, and consequently modifies the entire accessible energy region.
From eq. \eqref{56.1}, the inverse of the temperature in terms of energy for the NC LW potential is
\begin{equation}
    \label{57}
    \frac{1}{T_{NCLWP}(E)} = \frac{1}{g_{NCLWP}(E)}\frac{\partial g_{NCLWP}(E)}{\partial E} = \frac{1}{g_{NCLWP}{(E)}}\left\{\frac{\partial g_{LWP}^{(0)}(E)}{\partial E} + \Theta\frac{\partial g_{LWP}^{(1)}(E)}{\partial E} \right\} \; ,
\end{equation}
where
\begin{eqnarray}
\label{58}
\frac{1}{T_{NCLWP}(E)}&=& \frac{1}{g_{NCLWP}{(E)}}\Bigg\{    \frac{g_{LW}^{(0)}(E)}{E}- \frac{2 A R^{3} k^{2} E}{\mu }\Bigg\{\frac{2}{E} A_{1}(E) + \frac{1}{\mu k r_{\max}} A_{2}(E) \nonumber\\       &-&  \frac{\mu}{2 E^{2}} \left[ \left( r_{\max}  - \frac{E}{\mu^{2} r_{\max}} \right)  \left(1 - \frac{E r_{\max}}{k}\right)^{2}- b \left(      1 - \frac{E b}{k} \right)^{2} \right]  \nonumber \\  
&+&   \frac{e^{-\mu r_{\max}}}{E^{2}} \left[ 1 - \frac{e^{-\mu r_{\max}}}{4} + \frac{E}{4 \mu k r_{\max}} e^{-\mu r_{\max}} - \frac{E^{2}}{\mu^{2} k^{2} r_{\max}} \right]  \nonumber\\  
&+& \frac{e^{-\mu b}}{E^{2}} \left( 1 - \frac{e^{-\mu b}}{4}\right)  \Bigg\} + 2AR^3 k M\Theta\Bigg\{ \frac{kE}{\mu} \Bigg[ \frac{\mu}{2E}      \left( 1 - \frac{E r_{\max}}{k} \right)^{2}  \nonumber \\  &-&  \mu e^{- \mu r_{\max}} \Bigg[      \frac{1}{E} \left(  1 - \frac{e^{-\mu r_{\max}}}{4} \right) - \frac{1}{k}  \left( r_{\max} + \frac{1}{\mu} \right)\Bigg] \nonumber \\  
&+&  e^{-\mu r_{\max}} \left( \frac{\mu}{4E} e^{-\mu r_{\max}} - \frac{1}{k} \right) \Bigg]A_3(E) -\Bigg[ \log\left( \frac{r_{\max}}{b} \right) - \mbox{Ei}(-\mu r_{\max})  \nonumber\\ 
&+&  \mbox{Ei}(-\mu b) \Bigg] -\Bigg[ E  \, \frac{1 - e^{-\mu r_{\max}}}{r_{\max}}    + k \Bigg( -\frac{(1 - e^{-\mu r_{\max}})^2}{r_{\max}^2}   \nonumber\\
    &+&  \frac{4 \mu \left( e^{-\mu r_{\max}} - e^{-2 \mu r_{\max}} \right)}{r_{\max}} \Bigg) \Bigg] \frac{\sqrt{2}}{2\mu k}\sqrt{\frac{k}{\mu k - E}}  \Bigg\} \Bigg\} \; ,
\end{eqnarray}
where
\begin{subequations}
\begin{eqnarray}
    \label{57.1}
        A_1(E) &=& \frac{\mu K}{6 E^2} \left[ \left( 1 - \frac{E r_{\max}}{K} \right)^3 - \left( 1 - \frac{E b}{K} \right)^3 \right]  \; , \\
        A_2(E) &=& - e^{-\mu r_{\max}} \left[ \frac{1}{E} \left( 1 - \frac{e^{-\mu r_{\max}}}{4} \right) - \frac{1}{K} \left( r_{\max} + \frac{1}{\mu} \right) \right] \; ,\\
        A_3(E) &=&\frac{M \mu}{k\sqrt{\mu - \dfrac{E}{k}}
    \left( \dfrac{\sqrt{2}}{\mu} \sqrt{\mu - \dfrac{E}{k}} - \dfrac{3}{2} \right)} + \frac{\left( 1 - \frac{1}{2} \sqrt{\mu - \dfrac{E}{k}} \right)}{2\left(\mu - \dfrac{E}{k}\right)
    \left( \dfrac{\sqrt{2}}{\mu} \sqrt{\mu - \dfrac{E}{k}} - \dfrac{3}{2} \right)} \; .
\end{eqnarray}
\end{subequations}
This result shows that the corrections introduced by the noncommutativity modify the limits of integration, and introduce quantum spatial-structure effects into the thermal behavior of the system. For small energy range of $E$, the effects of $\Theta$ may dominate the thermal dynamics. In higher energy regimes, the exponential terms decay rapidly, and the system tends to exhibit classical behavior. The usual result presented in \cite{15} is recovered when $\Theta =0$.
%
%

%
 \section{The non-commutative approach in the statistical mechanics of Boltzmann-Gibbs }
 To obtain the thermal capacity of these potentials written in the context of NC theory, we can use the statistical mechanics of Boltzmann-Gibbs. So, for the calculation of the heat capacity, we need to obtain the average energy of the system, which is related to the partition function. The general form for the partition function of a system of two particles that interact via potential $V(r)$ has the form
\begin{eqnarray}\label{65}
Z_{D}=\int_{-\infty}^{\infty} e^{-\beta H} \, d^{D}x \, d^{D}p \; ,
 \end{eqnarray}
 where $\beta = T^{-1}$ (considering the Boltzmann constant $\kappa_{B} = 1$), $H$ is the Hamiltonian of the system, and $D$ is the number of dimensions in the space. As in the calculation of the density of states, we assume that the radial distance $(r)$ fluctuates in the interval $(b, R)$ in the calculation of the partition function, mean energy and heat capacity.

The thermodynamic quantities are derived within a semiclassical framework, 
combined with a perturbative expansion in both the NC parameter $\Theta$ and 
the Boltzmann factor. It is therefore essential to clearly specify the regime of validity 
of these approximations.

The NC corrections are treated up to first order in $\Theta$. This approximation 
is valid under the condition that the characteristic NC scale is small compared 
to the typical length scales of the system. More precisely, one requires
\begin{equation}
\frac{\Theta M}{r^2} \ll 1 \; ,
\end{equation}
where $M$ is the conserved quantity defined in Eq.~(8), and $r$ represents the relevant 
radial scale of the interaction. The factor $\frac{\Theta M}{r^2}$ is the non-commutative correction relative to the principal term of the potential, for both the Yukawa and Lee-Wick potentials, and it measures how important the non-commutative part is compared to the usual potential.
 This ensures that higher-order contributions in $\Theta$ 
remain subleading. 

That way, in the canonical ensemble, the partition function is computed assuming
\begin{equation}
|\beta V(r)| \ll 1 \; ,
\end{equation}
which allows us to expand the exponential $e^{-\beta V(r)}$ perturbatively. This condition 
is satisfied in the high-temperature regime (small $\beta$), or when the interaction 
potential is sufficiently weak within the integration domain $r \in (b, R)$.

For the Yukawa potential, this condition is typically well controlled due to its exponential 
decay. However, for the NC Lee--Wick potential, the situation is more subtle. 
Although the commutative Lee--Wick potential is finite at the origin, the NC 
correction introduces a term behaving as $r^{-3}$, which can enhance the magnitude of 
$\beta V(r)$ near the lower cutoff $r = b$.
Therefore, the perturbative expansion is reliable only if
\begin{equation}
\beta \, \max_{r \in (b,R)} |V(r)| \ll 1 \; ,
\end{equation}
which imposes a combined constraint on temperature, cutoff scale $b$, and the strength 
of the interaction.

Furthermore, the statistical treatment assumes a semiclassical regime in which the phase space is 
treated as continuous. This requires that the typical action scale of the system is large 
compared to $\hbar$ (here set to unity), and that quantum interference effects can be 
neglected. In this sense, the NC parameter $\Theta$ introduces a deformation 
of the classical phase space, but does not fully capture quantum corrections beyond the 
semiclassical level.

\subsection{Heat capacity for the NC Yukawa potential}


We substitute the Hamiltonian of the two-particle interacting through the NC Yukawa potential eq.(\ref{2.9}) in the eq. (\ref{65}),
that yields the partition function
\begin{eqnarray}\label{68}
Z_{NCYP}(\beta)=4\pi\left(\frac{2\pi\mu}{\beta}\right)^{3/2}\int_b^R e^{-\beta V_{\text{NCYP}}(r)} \, r^2 \, dr \; ,
 \end{eqnarray}
and, taking into account the condition $|\beta V(r)| \ll 1$, we can write eq.(\ref{68}) as
 \begin{eqnarray}\label{70}
Z_{NCYP}(\beta) &=& 4\pi\left(\frac{2\pi\mu}{\beta}\right)^{3/2}\int_b^R r^2 \, \left[ 1 + \beta\left(\frac{k e^{-\mu r}}{r} + \frac{\Theta k M e^{-\mu r}}{r^3}\right) 
\right. 
\nonumber \\ 
&&
\left. 
+ \frac{\beta^2}{2}\left(\frac{k^2 e^{-2\mu r}}{r^2} + \frac{2\Theta k^2 M e^{-2\mu r}}{r^4}\right) \right] \; ,
\end{eqnarray}
keeping terms up to first order in $\Theta$. By integrating term by term, we obtain the final form of the partition function given by
\begin{equation}\label{70.1}
Z_{NCYP}(\beta) = 4\pi \left(\frac{2\pi\mu}{\beta}\right)^{3/2} \left[ \, I_0 + \beta \, k \, A_1 + \beta \, \Theta \, k \, M \, A_2 + \frac{1}{4} \, \beta^2 \, k^2 \, A_3 + \Theta \, \beta^2 \, k^2 \, M \, A_4 \, \right] \; ,
\end{equation}
where we have defined the constants $A_{i} \,(i=1,2,3,4)$ and $I_{0}$ as
\begin{subequations}
\begin{align}\label{70.2}
A_1 &= \frac{1}{\mu^2} \left[ \, e^{-\mu b} \, (1 + \mu b) - e^{-\mu R} \, (1 + \mu R) \, \right] \; ,  \\
A_2 &= \mbox{Ei}(\mu b) - \mbox{Ei}(\mu R) \; , \\
A_3 &= \frac{e^{-2\mu b} - e^{-2\mu R}}{\mu} \; ,  \\
A_4 &= \frac{e^{-2\mu b}}{b} - \frac{e^{-2\mu R}}{R} + 2\mu\left[\,\mbox{Ei}(2\mu b) - \mbox{Ei}(2\mu R)\,\right] \; ,  \\
I_0 &= \frac{R^3 - b^3}{3} \; .
\end{align}
\end{subequations}
%
%

\begin{figure}
    \centering
    \includegraphics[width=0.8\linewidth]{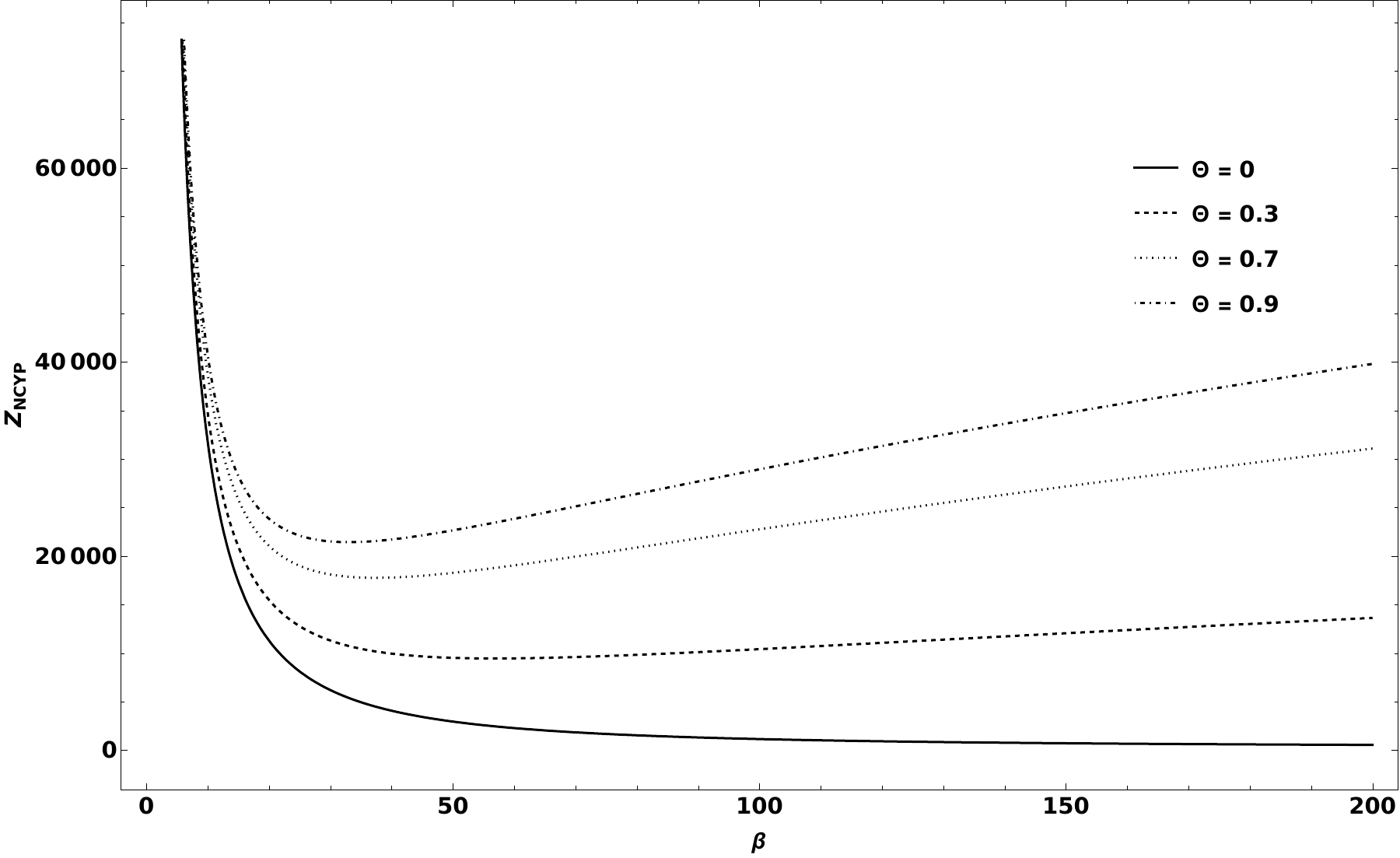}
    \caption{The partition function for the NC Yukawa potential as function of $\beta$ for $\Theta=0$, $\Theta=0.3$, $\Theta=0.7$, and $\Theta=0.9$ in area unities.}
    \label{fig1}
\end{figure}

\begin{figure}
\centering
\includegraphics[width=0.8\linewidth]{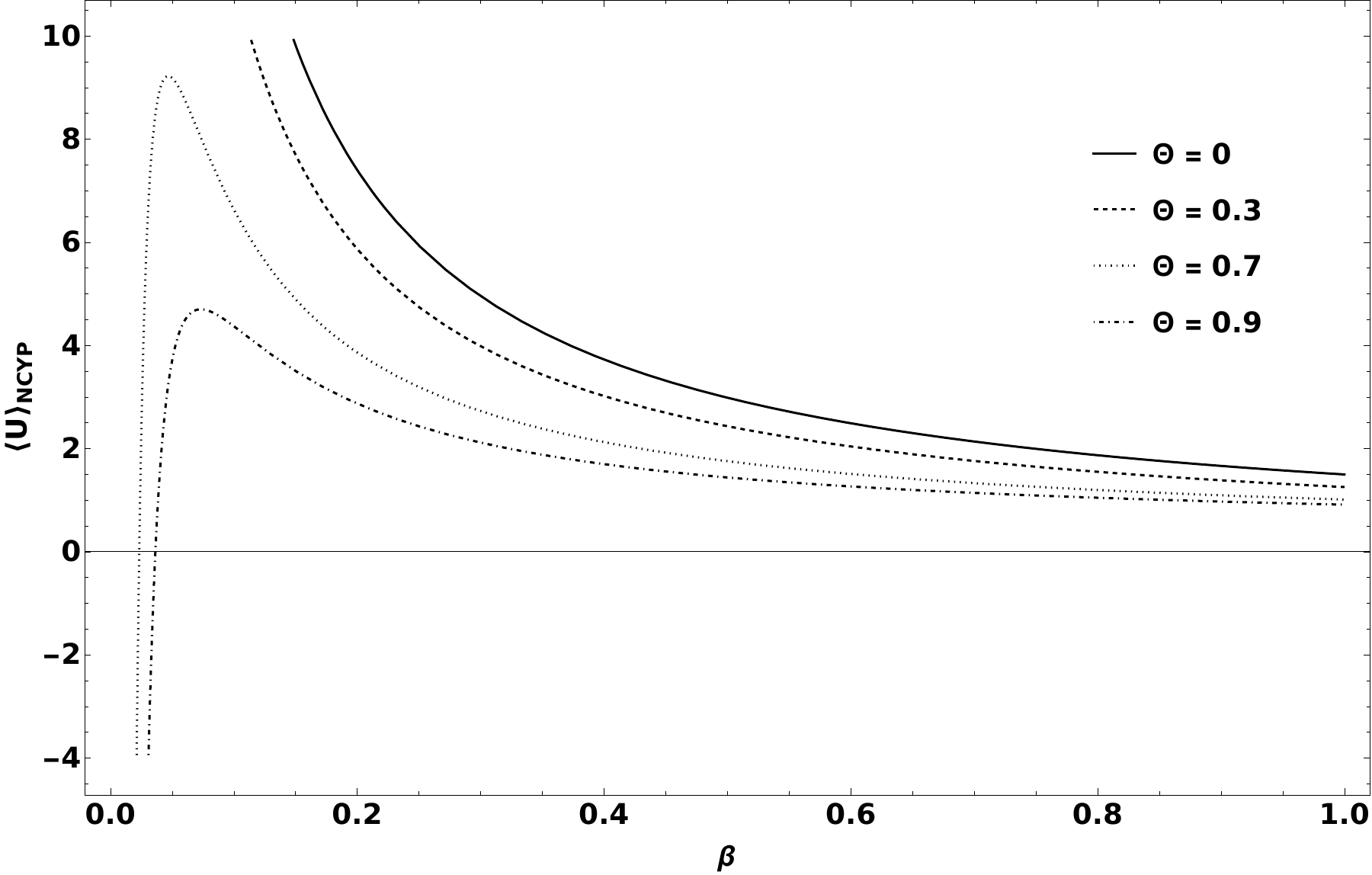}
\caption{The mean energy for the NC Yukawa potential as function of $\beta$. We use $\Theta=0$, $\Theta=0.3$, $\Theta=0.7$, and $\Theta=0.9$ in this plot in area unities.}
\label{fig2}
\end{figure}

%
We plot the partition function eq.(\ref{70.1}) as a function of $\beta$ in figure \ref{fig1} for the values of $\Theta=0$, $\Theta=0.3$, $\Theta=0.7$, and $\Theta=0.9$. As we can see in figure \ref{fig1} at high temperatures (when $\beta \rightarrow 0$), the contributions due to the NC $\Theta$-parameter are very small, but they grow significantly at low temperatures ($\beta \rightarrow \infty$).
With the help of the partition function, one can determine the thermodynamic quantities, as the mean energy and heat capacity for a system ruled by the hamiltonian function.
%
%
We start by computing the mean energy associated with the NC Yukawa partition function:
 \begin{eqnarray}\label{71}
\left<U\right>_{NCYP}(\beta)&=&  -\frac{\partial}{\partial \beta} \ln Z_{NCYP}(\beta) \nonumber \\ &=&  \frac{3}{2\beta} - \frac{1}{\beta^2} \frac{k A_1 + \Theta k M A_2 + \frac{1}{2}\beta k^2 A_3 + 2\beta \Theta k^2 M A_4}{I_0 + \beta k A_1 + \beta \Theta k M A_2 + \frac{1}{4}\beta^2 k^2 A_3 + \beta^2 \Theta k^2 M A_4} \; ,
\end{eqnarray}
 in which the constants $A_{i}\,(i=1,2,3,4)$ are defined in the eq. (\ref{70.2}). We plot the mean energy eq. (\ref{71}) as a function of $\beta$ in the figure \ref{fig2} using the values of $\Theta=0$, $\Theta=0.3$, $\Theta=0.7$, and $\Theta=0.9$. The mean energy curve increases as $\beta$ increases until it reaches a maximum value, and then decreases. Notice that the smaller the correction due to the NC $\Theta$-parameter, the higher the peak at high temperatures.
%

%

The heat capacity of the NC Yukawa potential has the result
\begin{eqnarray}\label{72}
C^{NCYP}_{V}(\beta) = -\beta^2 \, \frac{\partial \langle U \rangle}{\partial \beta} 
=\frac{3}{2} - \frac{2N}{\beta D} + \frac{N^{\prime} \, D - N\,D^{\prime}}{D^2} \; ,
\end{eqnarray}
in which the constants $N$, $D$, $N^{\prime}$ and $D^{\prime}$ are defined by
\begin{subequations}
\begin{align}
\label{72.1}
N &= k \, A_1 + \Theta \, k \, M \, A_2 
+ \frac{1}{2}\, \beta \, k^2 \, A_3 + 2 \Theta \, \beta \, k^2 \, M \, A_4 \; ,  \\
D &= I_0 + \beta \, k \, A_1 + \Theta \, \beta \, k \, M \, A_2 + \frac{1}{4} \, \beta^2 \, k^2 \, A_3 + \beta^2 \, \Theta \, k^2 \, M \, A_4 \; , 
\\
N^{\prime} &= \frac{1}{2} \, k^2 \, A_3 + 2\Theta \, k^2 \, M \, A_4 \; , \\
D^{\prime} &= k \, A_1 + \Theta \, k \, M \, A_2 + \frac{1}{2}\, \beta \, k^2 \, A_3 + 2 \Theta \, \beta \, k^2 \, M \, A_4 \; .
\end{align}
\end{subequations}
\begin{figure}
    \centering
    \includegraphics[width=0.8\linewidth]{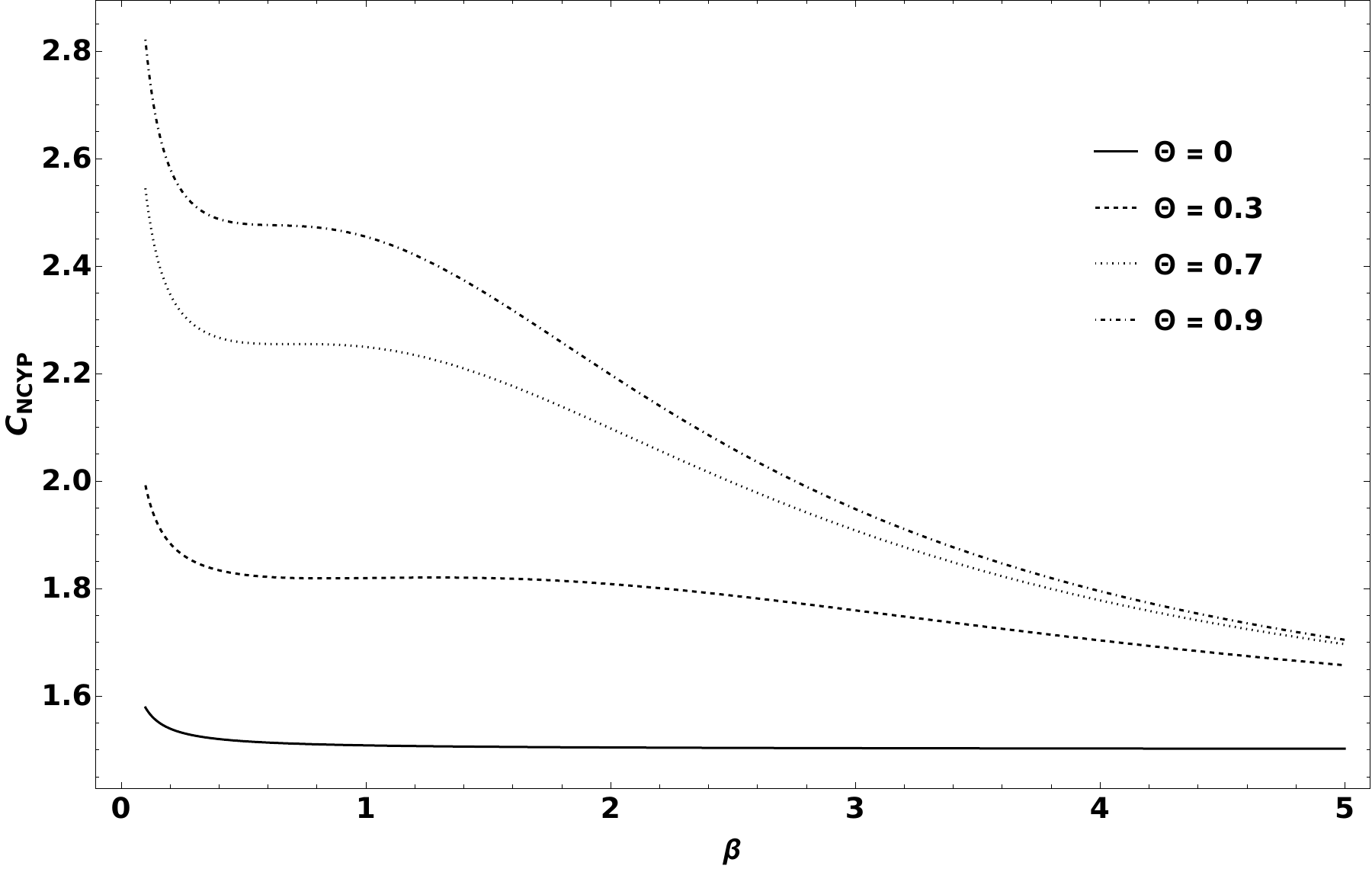}
    \caption{The heat capacity for the NC Yukawa potential as function of $\beta$ for the values of $\Theta=0$, $\Theta=0.3$, $\Theta=0.7$, and $\Theta=0.9$ in area unities.}
    \label{fig3}
\end{figure}

The heat capacity eq.(\ref{72}) as a function of $\beta$ is illustrated in figure \ref{fig3}. The values of the $\Theta$-parameter are $\Theta=0$, $\Theta=0.3$, $\Theta=0.7$, and $\Theta=0.9$ in this plot.We observe that the heat capacity $C_V^{NCYP}$ decreases monotonically with increasing $\beta$. This behavior becomes more pronounced as the NC parameter $\Theta$ increases, indicating that phase-space deformation enhances thermal sensitivity in this regime. Notice that the decrease is faster for larger values of the NC $\Theta$-parameter.
%

%

\subsection{Heat capacity for the NC Lee-Wick potential}
We substitute the Hamiltonian of the two-particle interacting through the NC Lee-Wick potential eq.(\ref{2.14}) in eq. (\ref{65}) to obtain the radial integral :
\begin{eqnarray}
Z_{NCLWP}(\beta)=4\pi\left(\frac{2\pi\mu}{\beta}\right)^{3/2}\int_b^R e^{-\beta\, V_{\text{NCLWP}}(r)} \; r^2 \; dr \; .
 \end{eqnarray}
Considering the Lee–Wick potential under the condition $|\beta\, V(r)| \ll 1$, we derive the correspondent partition function in first order in the $\Theta$-parameter :
\begin{eqnarray}\label{73}
Z_{NCLWP}(\beta) &=& 4\pi\left(\frac{2\pi\mu}{\beta}\right)^{3/2}\int_b^R r^2 \, \left[ \, 1 - \beta \, \frac{k}{r}\left(1 - e^{-\mu r}\right) 
- \beta \, \Theta \, \frac{M k}{r^3} \left(1 - e^{-\mu r}\right) 
\right. 
\nonumber \\ 
&&
\left.
+ \beta^2 \, \frac{k^2}{2r^3} \left(1 - e^{-\mu r}\right)^2 + \beta^2 \, \Theta \, \frac{k^2 M}{r^4}\left(1 - e^{-\mu r}\right)^2 \, \right] dr \; .
 \end{eqnarray}
After the radial integrations, the partition function is
\begin{equation}\label{73.1}
Z_{NCLWP}(\beta) = \frac{4\pi V}{h^3} \left( \frac{2\pi m}{\beta} \right)^{3/2} \left[ \, I_0 - \beta \, I_1^{(0)} + \frac{1}{2} \, \beta^2 \, I_2^{(0)} - \beta \, \Theta I_1^{(1)} + \beta^2 \, \Theta \, I_2^{(1)} \, \right] \; ,
\end{equation}
where the constants are defined by
\begin{subequations}
\begin{eqnarray} \label{73.2}
I_0 &=& \frac{R^3 - b^3}{3} \; ,
\\
I_1^{(0)} &=& k \left[ \frac{R^2 - b^2}{2} + \frac{e^{-\mu R}(1 + \mu R) - e^{-\mu b}(1 + \mu b)}{\mu^2} \right] \; , 
\\
I_1^{(1)} &=& M k \left[ \ln\frac{R}{b} - \text{Ei}(-\mu R) + \text{Ei}(-\mu b) \right] \; ,
\\
I_2^{(0)} &=& k^2 \left[ (R - b) + \frac{2}{\mu}(e^{-\mu R} - e^{-\mu b}) - \frac{1}{2\mu}(e^{-2\mu R} - e^{-2\mu b}) \right] \; ,  \\
I_2^{(1)} &=& M k^2 \left[
\frac{1 - 2e^{-\mu b} + e^{-2\mu b}}{b} - \frac{1 - 2e^{-\mu R} + e^{-2\mu R}}{R} \right]
\nonumber\\
&&
+ 2\mu Mk^2 \left[ \, \text{Ei}(-2\mu b) - \text{Ei}(-2\mu R)
-\text{Ei}(-\mu b) + \text{Ei}(-\mu R) \, \right] \; .
\end{eqnarray}
\end{subequations}
\begin{figure}
    \centering
    \includegraphics[width=0.8\linewidth]{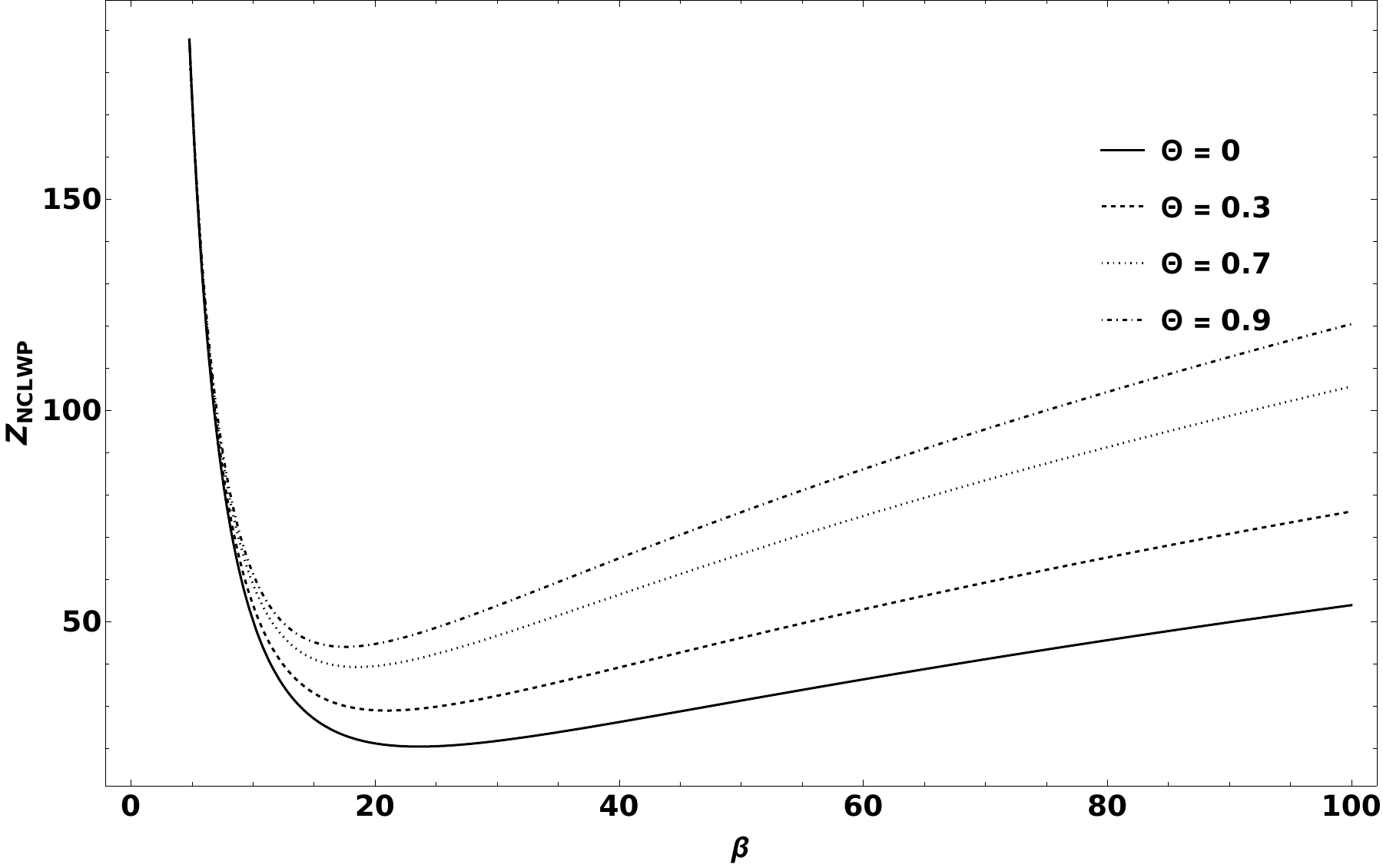}
    \caption{The partition function for the NC Lee-Wick potential as function of $\beta$, with the values of $\Theta=0$, $\Theta=0.3$, $\Theta=0.7$ and $\Theta=0.9$ in area unities.}
    \label{fig4}
\end{figure}
We plot the partition function eq.(\ref{73.1}) as a function of $\beta$ in figure \ref{fig4}. As we can see in figure \ref{fig4}, the contributions due to the NC $\Theta$-parameter are very small at high temperatures ($\beta \rightarrow 0$). They grow significantly at low temperatures ($\beta \rightarrow \infty$).
Therefore, we can calculate the mean energy and the heat capacity using the partition function. The mean energy as function of the temperature is
%
%
%
\begin{eqnarray}\label{74}
\langle U \rangle_{NCLWP} &=&
-\frac{\partial}{\partial \beta} \ln Z_{NCLWP}(\beta) 
\nonumber \\ 
&=&  
\frac{3}{2\beta} + \frac{I_1^{(0)} - \beta I_2^{(0)}}{D_0(\beta)} + \Theta \left[ \frac{I_1^{(1)} - 2\beta I_2^{(1)}}{D_0(\beta)} + \frac{\beta(I_1^{(0)} - \beta I_2^{(0)})(I_1^{(1)} - \beta I_2^{(1)})}{D_0(\beta)^2}\right] \; , \; \; \;\;\;\;
\end{eqnarray}
 where 
 \begin{equation} \label{74.1}
D_0(\beta) = I_0 - \beta \, I_1^{(0)} + \frac{1}{2} \, \beta^2 \, I_2^{(0)} \; .
 \end{equation}
 \begin{figure}
     \centering
     \includegraphics[width=0.8\linewidth]{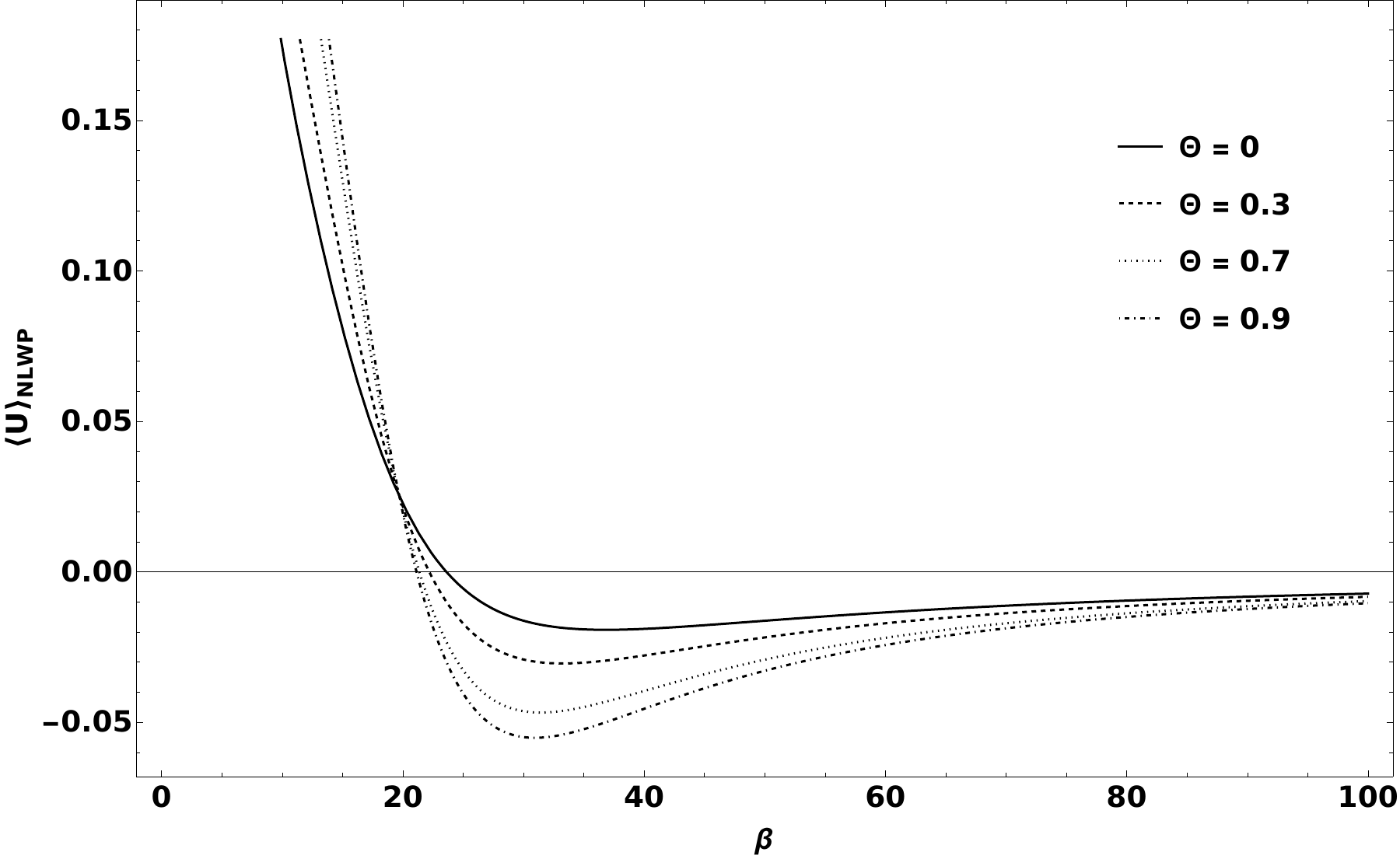}
     \caption{The nean energy for the NC Lee-Wick potential versus the inverse of the temperature $(\beta)$ for $\Theta=0$, $\Theta=0.3$, $\Theta=0.7$ and $\Theta=0.9$ in area unities.}
    \label{fig5}
 \end{figure}
The mean energy eq.(\ref{74}) versus the inverse of the temperature $(\beta)$ is illustrated in figure \ref{fig5}. We observed that the greater the $\Theta$-parameter, the higher the initial rise of the curve at high temperatures, and it decreases with $(\beta)$ until it reaches the minimum that represents a regime, where the system is more strongly coupled due to the LW potential. The non-commutativity intensifies this coupling : greater depth of the well. All slowly converge to values close to each other as $\beta$ increases (or when the temperature is decreasing). 
%

%
%

The heat capacity for the NC LW potential is
\begin{eqnarray} \label{74.2}
C_{V}^{NCLW} \!&=&\!  -\beta^2 \, \frac{\partial \langle U \rangle}{\partial \beta} = \frac{3}{2} + \beta^2 \left[ \, \frac{I_2^{(0)} D_0(\beta) - (I_1^{(0)} - \beta \, I_2^{(0)})^2}{D_0^2(\beta)} \, \right] 
\nonumber \\
&&
\hspace{-0.5cm}
+\Theta\,\beta^2 \left[ \, \frac{2I_2^{(1)} D_0(\beta) + (I_1^{(1)} - 2\beta I_2^{(1)}) D_0'(\beta)}{D_0^2(\beta)} - \frac{B'(\beta) D_0(\beta) - 2B D_0'(\beta)}{D_0^3(\beta)} \, \right] \; ,
\end{eqnarray}
 where
\begin{subequations}
\begin{eqnarray}
D_0^{\prime}(\beta) &=& -I_1^{(0)} + \beta I_2^{(0)} \; ,
\\
B^{\prime}(\beta) &=& (I_1^{(0)} - \beta I_2^{(0)})(I_1^{(1)} 
- \beta \, I_2^{(1)}) - \beta \, \left[ \, I_2^{(0)}(I_1^{(1)} - \beta I_2^{(1)}) + I_2^{(1)}(I_1^{(0)} - \beta \, I_2^{(0)}) \, \right] \; . \; \; \;
 \end{eqnarray}
 \end{subequations}
\begin{figure}
    \centering
    \includegraphics[width=0.8\linewidth]{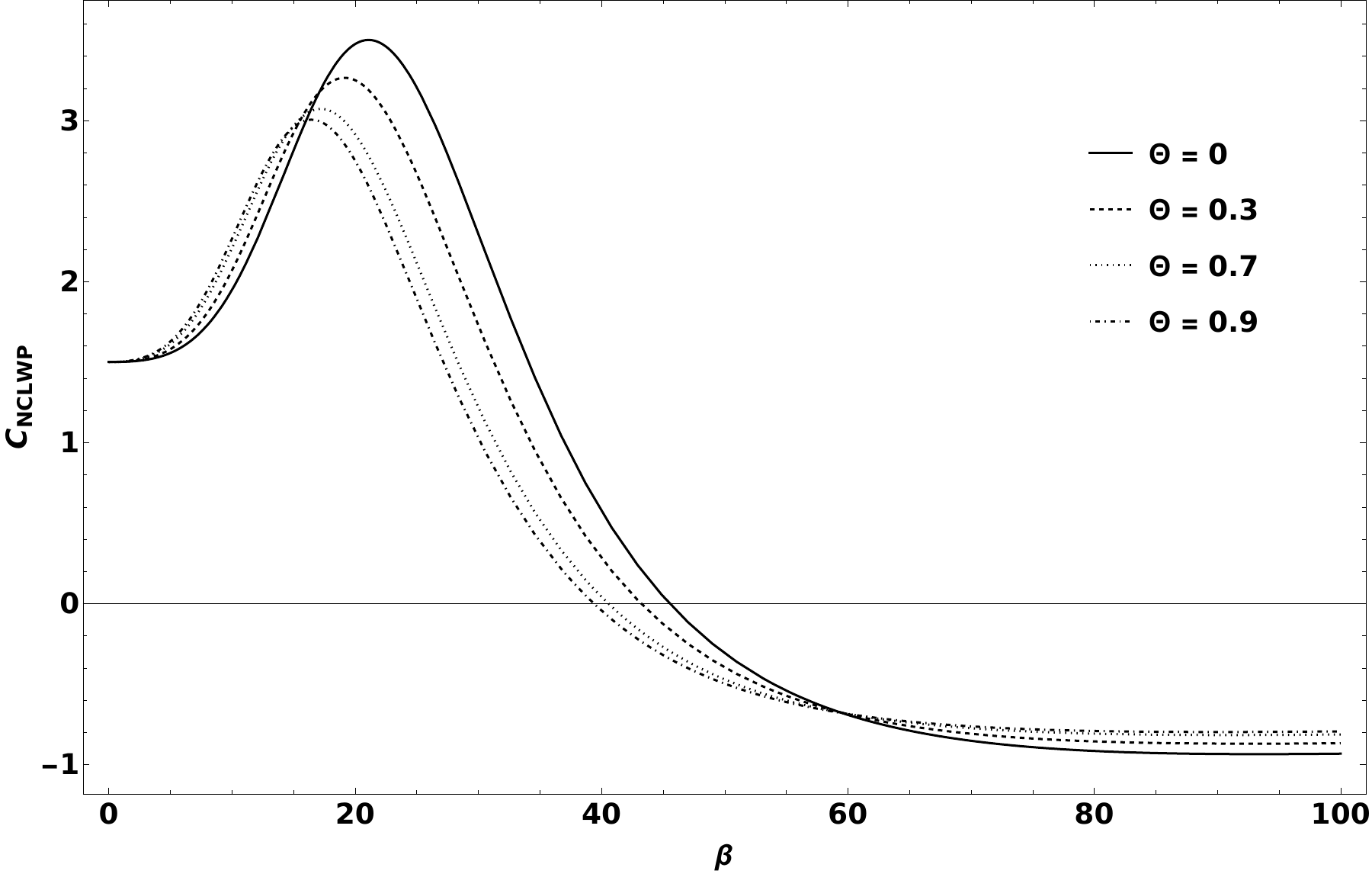}
    \caption{The heat capacity for the NC Lee-Wick potential as function of $(\beta)$ for the values of $\Theta=0$, $\Theta=0.3$, $\Theta=0.7$ and $\Theta=0.9$ in area unities.}
    \label{fig6}
\end{figure}
We plot the heat capacity eq.(\ref{74.2}) versus the inverse of the temperature $(\beta)$ in figure \ref{fig6}. We observe that the curves (with the values of $\Theta$-parameter) for the NC Lee-Wick potential reveal the presence of the term proportional to $\Theta$ plays a role in modifying the thermal response on the system. While its influence is minimal at high temperatures, it becomes dominant in the intermediate regime, resulting in higher and shifted peaks, intensifying the thermal dynamics. At low temperatures, the noncommutativity reinforces the effects of the perturbative approximation, resulting in negative thermal capacity within the truncated model.

We emphasize that the emergence of negative heat capacity occurs in a regime where the condition $|\beta V(r)| \ll 1$ may be marginally violated near the cutoff $r=b$, particularly due to the $r^{-3}$ NC contribution. Therefore, the result should be interpreted as indicative of a thermodynamic instability rather than a quantitatively exact prediction.

While the NC Yukawa system exhibits smooth thermodynamic corrections, the NC Lee–Wick system develops strong short-distance enhancements that qualitatively modify its thermodynamic stability.

The appearance of negative thermal capacities in the present model requires careful physical interpretation. It is well known that negative thermal capacities do not necessarily constitute a problem, but can naturally arise in specific classes of systems. 

An example is provided by self-gravitating systems, where the total energy is negative and the temperature increases as energy is removed, leading to $C < 0 $ \cite{paper1,paper2}. In these systems, the long-range nature of the interaction leads to the breakdown of standard thermodynamic assumptions, including the equivalence between statistical ensembles.

More recently, negative thermal capacities have also been discussed in a broader context, including systems with effective interactions and non-trivial phase space structures. In particular, the works of \cite{paper3,paper4} show that thermodynamic anomalies, such as negative heat capacity, can arise in systems where the underlying dynamics or geometry induces strong correlations or non-standard energy distributions.

More generally, systems with long-range interactions or strong correlations may exhibit thermodynamic anomalies, such as negative heat capacity, especially in the microcanonical ensemble \cite{paper5}. These features are often associated with phase transitions, metastable states, or dynamic instabilities. 

In the present case, although the Lee-Wick potential is intrinsically short-range, the NC correction introduces a term proportional to $r^{-3}$, which significantly enhances the interaction at short distances. This contribution modifies the effective structure of phase space and may induce behaviors resembling those of systems with long-range interactions or strong correlations.

From this perspective, the emergence of negative heat capacity can be interpreted as a manifestation of the sensitivity of thermodynamic quantities to the implicit geometry of phase space, as well as to the presence of strong correlations at short distances. At the same time, it is important to emphasize that this result is obtained within a perturbative framework based on the condition  $|\beta V(r)| \ll 1$.

The concept of negative heat capacities for both black holes and interacting classical systems like stars is nowadays well understood.   The physical structure of many well known thermodynamical frameworks is difficult in both cases because of the underlying role played by gravitation.   A system with negative heat capacity warms up by losing heat and cools down by gaining it. Hence, if heat flows from hot to cold then any temperature difference relative to the reservoir will produce heat flows that increase the difference, \cite{landsberg} and references therein.

Noncommutativity introduces a structural and deep modification of statistical mechanics and consequently in thermodynamics.   The main effect is that the phase space is not a classical one and this effect alters the counting of microstates as well as the partition function and all the relative thermodynamical quantities including the heat capacity which is important in our analysis here.   Since the NC algebra in Eq. (3) deforms the basic phase space volume it modifies the energy spectrum of the quantum systems.   Consequently these effects propagate directly towards thermodynamics considering that the partition function is modified since $E_n$ changes.   Besides, the state density $g(E)$ has corrections that depend on $\Theta$ and the phase space integrals now have corrections terms.

Noncommutativity does not generate negative heat capacity by itself, but modifies the phase-space structure in such a way that regions of thermodynamic instability may emerge or be enhanced.

Both the entropy and heat capacity are well known as being affected by any perturbation of the microstructure of the system. The noncommutativity can, in well defined circumstances, induce or preserve a negative heat capacity mainly in gravitational systems and, in our case, in effective models where noncommutativity acts as a geometrical deformation of the phase space.   In other words, the noncommutativity alters the thermodynamical structure in order to allow, preserve or even to remove regions where $C<0$.

Technically we can say that the $\Theta$-parameter introduces a new scale given by $M_{\theta} \sim (\sqrt{\Theta})^{-1}$ and the basic consequences are that for $M \gg M_\theta$ we have that $C \approx C_{classic} <0$; for $M \sim M_{\theta}$ the value of $C$ can diverge, change the signal or approaches to zero; and for $M < M_{\theta}$ the system can become stable.   Hence, for systems confined in NC spaces, the noncommutativity reduces the effective number of microstates, introduces a kind of geometric cutoff and transforms the entropy as a nonextensive function.   For $C=T\frac{\partial S}{\partial T}$, negative values of $C$ imply that $\frac{\partial^2 S}{\partial U^2} >0$, which is typical for nonextensive systems \cite{nss,bms,nicolini}.   So, it is important to understand that noncommutativity does not create a $C<0$ behavior but reorganize phase transitions and can induce a thermal instability in effective nonextensive systems.

\section{Conclusions}
\label{Resultados e discussão}
In this paper, we carried out a detailed analysis of the statistical thermodynamics of systems governed by the Yukawa and Lee--Wick potentials within a phase space endowed with a NC structure, as defined by the Eq.~(\ref{relxipi}). This deformation modifies the underlying symplectic geometry, leading to corrections in the equations of motion that directly affect the microscopic dynamics and, consequently, the macroscopic thermodynamic behavior.

We derived the NC extensions of both Yukawa and Lee--Wick potentials, and investigated their thermodynamic properties within the canonical and microcanonical ensembles. Our results show that the introduction of the NC parameter $(\Theta)$ leads to explicit corrections in statistical quantities as the density of states, partition function, mean energy, and heat capacity.

A key result of this work is the qualitative difference between the two systems. While the NC Yukawa potential produces smooth and controlled thermodynamic corrections, the NC Lee--Wick potential develops strong short-distance enhancements due to the contribution with $r^{-3}$ in the radial distance, which significantly modifies the thermodynamic stability of the system. In particular, regions of negative heat capacity emerge, signaling the presence of thermodynamic instabilities associated with the modified phase-space structure.

We emphasize, however, that noncommutativity does not generate negative heat capacity by itself, but rather reshapes the phase-space geometry in such a way that unstable thermodynamic regimes may emerge or be enhanced. Moreover, these features appear in a regime in which the perturbative conditions $|\beta V(r)| \ll 1$ and $\Theta M / r^2 \ll 1$ may be marginally satisfied, particularly near the short-distance cutoff. Therefore, the results should be interpreted as indicative of underlying instabilities rather than quantitatively exact predictions.

Although the present semiclassical and perturbative treatment provides important insights, a full non-perturbative analysis of the partition function would be necessary to establish the robustness of these effects. This constitutes a natural direction for a forthcoming project.

Overall, our results show that NC phase-space deformations introduce an effective geometric scale that can qualitatively alter thermodynamic behavior, especially in systems where the short-distance interactions are enhanced. The framework developed here can be extended to other interaction potentials, confined systems, or external fields, opening new perspectives for the investigation of NC effects in statistical mechanics and thermodynamics.

\acknowledgments

\noindent MGS thanks CAPES (Coordena\c{c}\~ao de Aperfei\c{c}oamento de Pessoal de N\'ivel Superior) for the financial support.

\end{document}